\documentclass[preprint,superscriptaddress,footinbib,showkeys,showpacs,tightenlines]{revtex4} 
\usepackage{graphicx}
\usepackage{epstopdf}
\usepackage{dcolumn}
\newcommand{\be}{\begin{equation}}
\newcommand{\ee}{\end{equation}}
\newcommand{\bee}{\begin{eqnarray}}
\newcommand{\eee}{\end{eqnarray}}
\newcommand{\eq}{\end{quote}}
\newcommand{\nn}{\nonumber}

\newcommand{\Slash}[1]{\ooalign{\hfil/\hfil\crcr$#1$}}

\def\gsim{\displaystyle\mathop{>}_{\sim}}
\def\lsim{\displaystyle\mathop{<}_{\sim}}
\begin{document}      
\preprint{PNU-NTG-15/2005}
\title{A new $N^*(1675)$ resonance in the $\gamma N\to\eta N$ reaction}
\author{Ki-Seok Choi}
\email{kschoi@rcnp.osaka-u.ac.jp}
\affiliation{Research Center for Nuclear Physics (RCNP), Ibaraki, Osaka
567-0047, Japan}
\affiliation{Department of
Physics and Nuclear physics \& Radiation Technology Institute (NuRI),
Pusan National University, Busan 609-735, Korea} 
\author{Seung-il Nam}
\email{sinam@pusan.ac.kr}
\affiliation{Department of
Physics and Nuclear physics \& Radiation Technology Institute (NuRI),
Pusan National University, Busan 609-735, Korea} 
\author{Atsushi Hosaka}
\email{hosaka@rcnp.osaka-u.ac.jp}
\affiliation{Research Center for Nuclear Physics (RCNP), Ibaraki, Osaka
567-0047, Japan}
\author{Hyun-Chul Kim}
\email{hchkim@pusan.ac.kr}
\affiliation{Department of
Physics and Nuclear physics \& Radiation Technology Institute (NuRI),
Pusan National University, Busan 609-735, Korea} 

\date{January, 2006}
\begin{abstract}
We study the $\eta$-photoproduction focusing on the new nucleon
resonance which was observed at $\sqrt{s}=1675$ MeV with a narrow
decay width ($\sim 10$ MeV) in the recent GRAAL experiment.  Using an
effective Lagrangian approach, we compute differential cross sections
for the $\eta$-photoproduction.  In addition to $N^*(1675)$, we employ
three other nucleon resonances, i.e. $N^*(1535)$, $N^*(1650)$ and 
$N^*(1710)$, and vector meson exchanges which are the most relevant
ones to this reaction process.  As a result, we can reproduce the
GRAAL data qualitatively well and observe obvious isospin asymmetry
between the transition magnetic moments of $N^*(1675)$: $\mu_{\gamma
  nn^*}\gg\mu_{\gamma pp^*}$. 
\end{abstract}
\maketitle
\section{Introduction}
After the first experimental observation of a signal of the
pentaquark baryon $\Theta^+$ by the LEPS collaboration at SPring-8
which was motivated by the theoretical predictions from the chiral soliton 
model ($\chi$SM)~\cite{Diakonov:1997mm}, we have experienced abundant
research activities in hadron physics.  However, we still have many 
unknowns about the $\Theta^+$ baryon, and there have been strong
criticisms against it.  The negative results of the recent CLAS
experiment deepened the question on its
existence~\cite{DeVita:2005CLAS}.   
 
In this recent unsettled situation for $\Theta^+$, it is
very natural that there have been new theoretical and experimental
efforts for $\Theta^+$.  Theoretically, for instance, higher spin
states of $\Theta^+$ were suggested in the constituent quark model,
lattice QCD and reaction
studies~\cite{Hosaka:2004bn,Ishii:2005vc,Lasscock:2005kx,Nam:2005jz}.
Experimentally, the LEPS collaboration reported a new signal for
$\Theta^+$~\cite{Hotta:2005rh}.  $I=1\ \Theta^{++}$ and
the charmed pentaquark were also reported by the STAR
collaboration~\cite{Huang:2005}.  In addition to them, a recent GRAAL 
experiment announced a new nucleon resonance with a seemingly narrow
decay width $\sim 10$ MeV and a mass $\sim 1675$ MeV in the 
$\eta$-photoproduction.  This new nucleon-like resonance, $N^*(1675)$, 
may be regarded as a non-strange pentaquark because of its narrow
decay width, which is assumed to be one of the significant features of
typical pentaquark baryons, though one should not exclude a
possibility that it might be a known one among existing resonances.
Furthermore, the value of its mass,
$1675$ MeV, is close to that obtained by the $\chi$SM ($1710$ MeV).

Among these new experimental results, we focus on the GRAAL experiment
of the $\eta$-photoproduction in the present work.  We note that the
reaction process $\gamma N\to \eta N$ has been explored already 
experimentally as well as
theoretically~\cite{Tryasuchev:2003st,Benmerrouche:1994uc,Rebreyend:2004in, 
Crede:2003ax}.  It has been known from these previous studies that the 
nucleon-resonance pole ($N^*$) and vector meson-exchange contributions
prevail over those of the background.  Especially, the contribution
from the $N^*(1535)$ is the most dominant one near the threshold
($\sqrt{s}\sim1490$ MeV) region.  However, we have still many
theoretical ambiguities to be solved concerning this reaction process.
For instance, the value of the coupling constant $g_{\eta NN}$ lies in
a wide range ($0\sim7$), depending on either theoretical models or on
experiments to estimate the strength.  (see
Refs.~\cite{Holinde:1992ui,Grein:1979nw,
Piekarewicz:1993ad,Benmerrouche:1991qx,Zhu:2000eh,Stoks:1999bz}).  

In addition to the known facts of the $\eta$-photoproduction mentioned
above, a new interesting feature was observed in the GRAAL experiment:
The $N^*(1675)$ is preferably excited on the neutron target by
photons.  It implies that large isospin asymmetry may 
exist in the electromagnetic transitions for $N^* (1675) \to N\gamma$,
since the strong coupling constants $g_{\eta NN^*}$ are independent
of the nucleon isospin.  It would be difficult to explain the two
experimental observations, the narrow peak and its strong isospin
dependence, in terms of the conventional knowledge of meson-baryon
interactions.  

Recently, the values of the magnetic transitions, $\mu_{\gamma N
N^*}$, were estimated within the framework of the chiral
quark-soliton model ($\chi$QSM)~\cite{Kim:2005gz} in which the
$N^*(1675)$ was assumed to be a member of the baryon antidecuplet.
Interestingly enough, the results showed obvious isospin asymmetry
between $\mu_{\gamma n n^*}$ and $\mu_{\gamma p p^*}$ in
their magnitudes, though they depended on the nucleon $\Sigma$-term
rather sensitively.  In fact, the magnetic transition $\mu_{\gamma
pp^*}$ vanishes completely as a consequence of SU(3) flavor
symmetry when $N^*(1675)$ is assumed to belong to the antidecuplet,
while the $\mu_{\gamma nn^*}$ remains finite.  The result of the
$\chi$QSM is thus understood as a general consequence of flavor SU(3)
symmetry and its breaking.

In the present work, we investigate the $\eta$-photoproduction via the
$\gamma N\to \eta N$ reaction process theoretically, including the
$N^*(1675)$ in the framework of the effective Lagrangian method at the
tree-level calculation with gauge-invariant form
factors~\cite{Davidson:2001qs,Ohta:1989ji} employed.  For the
numerical calculations, various unknown parameters are 
determined by existing experimental data.  We consider both positive
and negative parities of $N^*(1675)$, since we have no experimental
information of its parity.  Furthermore, we consider
both positive and negative anomalous magnetic transition moments 
$\mu_{\gamma NN^*}$, since the sign of the coupling constant
is unknown to date.  

We will show in the present work that strong isospin 
asymmetry, i.e. $\mu_{\gamma nn^*} \gg \mu_{\gamma p
  p^*}$, does really exist in reproducing the GRAAL data.  The
estimated values are consistent with those from the
$\chi$QSM~\cite{Kim:2005gz} and the phenomenological 
estimation~\cite{Azimov:2005jj}.  As a result,
we will see that it is reasonable to regard the newly
observed narrow resonance peak by the GRAAL experiment as a member of
pentaquark baryon antidecuplet.   

This paper is organized as follows: In Section II we
will define the effective Lagrangians and construct the invariant
amplitudes.  We will also discuss various ways of determining
parameters.  Section III will be devoted to the numerical
results for the possible parameter combinations.  In the final
section, we will summarize our work.   
\section{Formalism}
\begin{figure}[tbh]
\includegraphics[width=12cm]{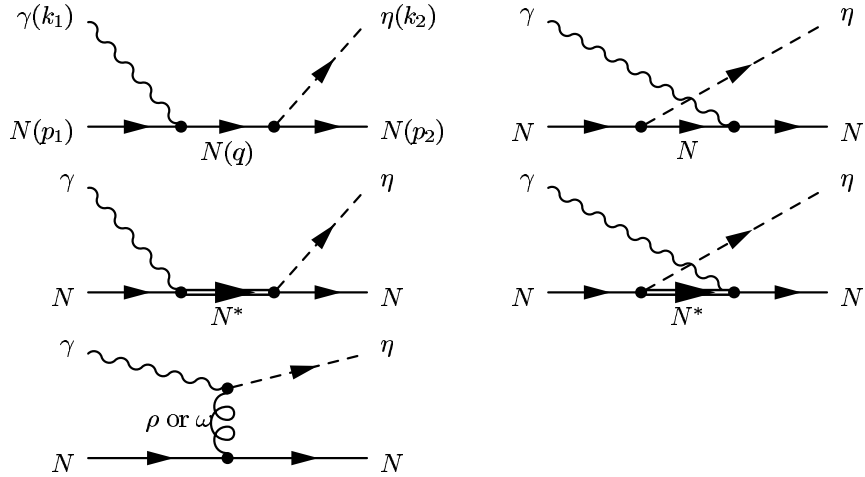}
\caption{Born diagrams calculated in the effective Lagrangian
  method. Top: nucleon pole contributions (left: $s$-channel and
  right: $u$-channel), Middle: nucleon-resonance pole contributions
  (left: $s$-channel and right: $u$-channel) and Bottom: the vector
  meson-exchange contributions ($t$-channel).}     
\label{fig0}
\end{figure}

We start with the effective Lagrangians for the interaction
vertices as depicted in Fig.~\ref{fig0}, where we also define the
four-momenta of scattering particles.  The effective Lagrangians are
given as 
\bee
 \mathcal{L}_{\gamma NN}
&=&-e\bar{N}\Slash{A}N - i
\frac{e\kappa_N}{2M_N}\bar{N}\sigma_{\mu\nu}q^{\nu}A^{\mu}N+h.c.,\nn\\
\mathcal{L}_{\eta NN}&=&-ig_{\eta NN}\bar{N}\gamma_5 \eta N+h.c.,\nn \\
\mathcal{L}_{VNN}&=& -g^v_{VNN} \bar{N}\Slash{V}N-
 i\frac{g^t_{VNN}}{2M_N}\bar{N}\sigma_{\mu\nu}q^{\nu}V^{\mu}N+h.c.,\nn\\
 \mathcal{L}_{\gamma N N^*}&=& \frac{e\mu_{\gamma N N^*}
 }{2(M_N+M_{N^*})}\bar{N}^*\Gamma_{\mu\nu}F^{\mu\nu}N+h.c.,\nn\\ 
 {\cal L}_{\eta N N^*}&=&-ig_{\eta N N^*} \bar{N }\Gamma \gamma_5 \eta
 N^*+h.c.,\nn \\ 
\mathcal{L}_{ \gamma\eta V}&=&\frac{eg_{ \gamma\eta
    V}}{4M_{\eta}}\epsilon_{\mu\nu\sigma\rho}F^{\mu\nu}V^{\sigma\rho}\eta+h.c.,   
\label{lagrangians}
\eee
where $N$, $N^*$, $\eta$ and $V$ stand for the fields corresponding to 
the nucleon, nucleon resonance, pseudoscalar meson $\eta$, and vector
mesons ($\rho$ and $\omega$), respectively.  With the parity of
the nucleon resonance $N^*$ distinguished, $\Gamma$ and
$\Gamma_{\mu\nu}$ are defined as follows;  
\bee
{\rm Positive\,\,parity}&:&\Gamma^{\mu\nu}=\sigma^{\mu\nu}\,\,\,\,{\rm
  and}\,\,\,\,\Gamma={\bf 1}_{4\times4},\nn\\ 
{\rm Negative\,\,parity}&:&\Gamma^{\mu\nu}=\gamma_5\sigma^{\mu\nu}
\,\,\,\,{\rm and}\,\,\,\,\Gamma=\gamma_5. 
\eee

Although there are about twenty nucleon resonances experimentally
known for the energy regions below $\sqrt{s}=2.0$ GeV, we only
consider four nucleon resonances: $N^*(1535,J^P=1/2^-)$,
$N^*(1650,J^P=1/2^-)$, $N^*(1675,J^P=$unknown$)$, and
$N^*(1710,J^P=1/2^+)$, which turn out to make major contributions to
the $\eta$-photoproduction.  We verified that other resonances gave
negligibly small contributions to the total amplitudes, especially
from the threshold to $E_{\rm CM} \leq 1.7$ GeV, which is the region
we are interested in. 

For the coupling constants of the nucleon, i.e. $g_{\eta N N}$,
$g_{\rho  NN}^{v,t}$ and $g_{\omega NN}^{v,t}$, we adopt the values
from the Nijmegen potential~\cite{Stoks:1999bz}, while the photon
couplings $g_{\rho \eta \gamma}$ and $g_{\omega \eta\gamma}$ are
determined by the radiative decays of $\rho$ and
$\omega$~\cite{Tiator:1994et} mesons.  Their values are listed in 
Table.~I.    
\begin{table}[h]
\begin{center}
\begin{tabular}{|c|c|c|c|c|c|c|}
\hline
$g_{\eta NN}$&$g_{\rho NN}^v$&$g_{\rho NN}^t$&
$g_{\omega NN}^v$&$g_{\omega NN}^t$&$g_{\rho \eta \gamma}$&$g_{\omega 
  \eta \gamma}$\\ 
\hline
0.47&2.97&12.52&10.36&4.20&0.89&0.192\\
\hline
\end{tabular}
\end{center}
\label{table0}
\caption{The relevant coupling constants used in the present work.  The
  meson-nucleon couplings are taken from the Nijmegen
  potential~\cite{Stoks:1999bz} and the
  meson-meson-photon ones from Ref.~\cite{Tiator:1994et}} 
\end{table}

In order to calculate the contributions of the nucleon-resonances, we
need to determine the resonance parameters $\kappa_{\gamma N N^*}$ and 
$g_{\eta N N^*}$.  For the known resonances $N^*(1535)$, $N^*(1650)$ and
$N^*(1710)$, we utilize the experimental data of the partial decay width
and electromagnetic helicity amplitudes~\cite{Eidelman:2004wy} via 
the following relations:
\bee
g_{\eta NN^*}&=&\sqrt{\frac{4\pi M_{N^*}\Gamma_{N* \to \eta
 N}}{|\vec{P}_f|M_N(\sqrt{1+\frac{|\vec{P}_f|^2}{M_N^2}} \pm 1)}},\nn\\  
|A_{\frac{1}{2}}|^2 &=& \left(\frac{e|\kappa_{\gamma
      NN^*}|}{M_{N^*}+M_N}\right)^2\frac{M^2_{N^*}-M^2_N}{2M_N}. 
\label{eq:exd}
\eee
In Eq.(\ref{eq:exd}), the $\pm$ sign corresponds to negative and 
positive parity resonances.  For the new resonance $N^*(1675)$, the
value of the total decay width, $\Gamma_{N^*(1675)\to all}$, was
estimated to be $\sim10$ MeV in the GRAAL
experiment~\cite{Kuznetsov:2004gy,Kuznetsov:2006de}.  We assume in the present
calculation that the decay process of $N^*\to \eta N$ is solely
explained by the total decay width from which the coupling constant
$g_{\eta NN^*(1675)}$ is determined.  We find $g_{\eta
NN^*(1675)}=2.8$ for the positive parity $N^*(1675)$, whereas $0.54$
for the negative parity one.  Furthermore, in order to make a better  
comparison with the experimental data which include the effect of the
Fermi motion of the neutron in the deuteron, we use an effective width
of $N^*(1675)$ $\Gamma_{\rm eff}=40$ MeV in the Breit-Wigner form.
The unknown is the electromagnetic coupling of $\mu_{\gamma N  
  N^*}$, which is the only parameter in our calculation.  We
vary the value of $\mu_{\gamma N N^*}$: $|\mu_{\gamma
NN^*}|\leq 0.3 \, \mu_N$~\cite{Kim:2005gz}.  The resulting coupling 
constants are listed in Table.~II.   
\begin{table}[t]
\begin{tabular}{|c|c|c|c|c|}
\hline
&$\Gamma_{N^*}$ [MeV]&$\Gamma_{N^*\to\eta N}/\Gamma_{N^*}$
[$\%$]&$A^n_{1/2}$ [GeV$^{-1/2}$]&$A^p_{1/2}$ [GeV$^{-1/2}$]\\ 
\hline
$N^*(1535)$&180 (150$\sim$200)&50
(30$\sim$55)&-0.065(-0.046$\pm$0.027)&0.087 (0.090$\pm$0.030)\\ 
$N^*(1650)$&150 (145$\sim$190)&7
(3$\sim$10)&-0.015(-0.015$\pm$0.021)&0.053 (0.053$\pm$0.016)\\ 
$N^*(1710)$&100 (50$\sim$250)&6 (6$\pm$1)&-0.002
(-0.002$\pm$0.014)&0.009 (0.009$\pm$0.022)\\ 
\hline
\end{tabular}
\caption{The parameters of the nucleon resonances: Full decay widths,
branching ratios and helicity amplitudes for the neutron and proton.} 
\label{table2}
\end{table} 

The invariant amplitudes are now given as follows:
\bee
i\mathcal{M}_s&=&\frac{eg_{\eta NN}}{\{(k_1+p_1)^2-M^2_N\}}
\bar{u}(p_2)[\gamma_5\{F^{N}_s\Slash{k}_1+F_c(\Slash{p}_1+M_N)\}
\Slash{\epsilon}\nn\\
&-&\frac{\kappa_NF^{N}_s}{2M_N}\gamma_5(\Slash{k}_1+\Slash{p}_1+M_N)
\Slash{\epsilon}\Slash{k}_1]u(p_1),\nn\\
i\mathcal{M}_u&=&\frac{eg_{\eta NN}}{\{(k_2-p_1)^2-M^2_N\}}
\bar{u}(p_2)[\Slash{\epsilon}\{F_c(\Slash{p}_2+M_N)-F^N_s
\Slash{k}_1)\gamma_5\nn\\&+&\frac{\kappa_NF^{N}_u}{2M_N}
\Slash{k}_1\Slash{\epsilon}(\Slash{p}_2-\Slash{k}_1+M_N)\gamma_5]u(p_1),\nn\\
i\mathcal{M}_t&=&\frac{-ieg_{\gamma\eta
    V}F^V_t}{M_{\eta}\{(k_1-k_2)^2-M^2_V\}}
\bar{u}(p_2)[g^v_{VNN}\epsilon_{\mu\nu\sigma\rho}k^{\mu}_1
\epsilon^{\nu}(k_1-k_2)^{\sigma}\gamma^{\rho}\nn\\
&+&\frac{g^t_{VNN}}{4M_N}\{\Slash{q}\epsilon_{\mu\nu\sigma\rho}
k^{\mu}_1\epsilon^{\nu}(k_1-k_2)^{\sigma}\gamma^{\rho}-
\epsilon_{\mu\nu\sigma\rho}k^{\mu}_1\epsilon^{\nu}(k_1-k_2)^{\sigma}
\gamma^{\rho}(\Slash{k}_1-\Slash{p}_1)\}]u(p_1)\nn\\
i\mathcal{M}_{s^*}&=&\frac{e\kappa_{\gamma N N^*} g_{\eta N
    N^*}}
{(M_N+M_{N^*})\{(k_1+p_1)^2-M^2_{N^*}-iM_{N^*}\Gamma_{\eta NN^*}\}}
\bar{u}(p_2)\gamma_5 \Gamma
(\Slash{k}_1+\Slash{p}_1+M_{N^*}) \Gamma\Slash{\epsilon}\Slash{k}_1u(p_1),\nn\\
i\mathcal{M}_{u^*}&=&\frac{e\kappa_{\gamma N N^*} g_{\eta N
    N^*}}{(M_N+M_{N^*})\{(k_2-p_1)^2-M^2_{N^*}-iM_{N^*}\Gamma_{\eta 
  NN^*}\}} \bar{u}(p_2)
\Gamma \Slash{\epsilon}\Slash{k}_1(\Slash{k}_2-\Slash{p}_1+M_{N^*})\gamma_5
\Gamma u(p_1).\nn\\
\label{amplitudes} 
\eee
The subscripts $s^*$ and $u^*$ denote the nucleon resonance pole terms
in the $s$- and $u$-channels, respectively, while the usual Born terms
of the nucleon are indicated by $s$, $u$ and $t$.  We note that the
nucleon resonance and vector meson pole terms are gauge-invariant.
We verified that the invariant amplitudes of Eq.~(\ref{amplitudes})
satisfied the Ward-Takahashi identity.  In order to take into account
extended structures of hadrons, we employ hadronic form factors which
preserve the Ward-Takahashi identity in terms of the prescription
proposed in Refs.~\cite{Davidson:2001qs,Ohta:1989ji}. They are 
parameterized as:
\bee
F^i_x=\frac{\Lambda^4}{\Lambda^4+(x-M^2_i)^2},
\label{formfactor}
\eee
where $x$ is the subscript indicating the Madelstam variables, $s$,
$t$, and $u$, while $i$ stands for the virtual particle in the
channel $x$.  We also employ one common form factor $F^N_c$ to make the
$s$- and $u$-channels satisfy the Ward-Takahashi identity: 
\bee
F_c=F^N_s+F^N_u-F^N_sF^N_u.
\label{commonformfactor}
\eee
This prescription of $F_c$ is determined by the 
normalization condition of the form factor, when exchange particles
are on mass-shell.  The cutoff parameter for the nucleon pole
terms is set to be $\Lambda \simeq 0.85$ GeV~\cite{Nam:2004xt}.  For
the resonance terms, we do not introduce the form factor as in
resonance dominant models~\cite{Benmerrouche:1994uc}.  For the
rho-meson exchange diagram, we, however, use a larger value of the
cut-off mass for the $\rho NN$
vertex~\cite{Benmerrouche:1994uc,Machleidt:1987hj}; $\Lambda \simeq
1.3$ GeV.        

\section{Numerical results}
As explained in the previous section, we consider several Born terms
including different nucleon resonances and vector meson
exchanges.  Theoretically, the relative signs of these terms are
important as they are added coherently in the amplitude.  We can
determine the signs of the resonance contributions of $N^*(1650)$ and
$N^*(1710)$ relative to the nucleon pole term through the signs of the
helicity amplitudes~\cite{Eidelman:2004wy}.  We do not know yet the
relative sign of the $\rho$- and $\omega$-exchange terms
and the new resonance term of $N^*(1675)$.  The relative sign of the
vector-meson terms is chosen in such a way that the total amplitude 
produces the observed energy dependence of the
$\eta$-photoproduction.  Note that due to the isospin structure
$\rho$-meson exchange changes the sign of the proton and neutron
targets, while $\omega$-meson exchange leaves it unchanged.  We also
considered the contribution of $N^*(1675)$ by changing the sign of the
electromagnetic coupling $\mu_{\gamma NN^*}$. 

Figures 2 and 3 draw our main results of the present investigation,
showing the energy dependence of the differential cross sections at  
$\theta=145$ degree.  These four panels in each figure correspond to
the cases of different isospin states($p$ or $n$) of $N^*(1675)$
targets with positive and negative parities.  The cross section are
computed by using different $\mu_{\gamma NN^*}$ and then are
compared with the data of the GRAAL experiment.  In all cases, the
thick lines represent the result without the $N^*(1675)$ contribution, 
where $\mu_{\gamma  NN^*}=0$.  In fact, the differential
cross sections were taken at several angles, $\theta=120 \sim 155$
degrees, within which the results are qualitatively similar.  

Before discussing the role of the $N^*(1675)$, we make general remarks. 
First, the largest peak around $E_{\rm CM}\sim 1530$ MeV is nicely
reproduced by choosing reasonable parameters for the $N(1535)$. 
Furthermore, the results are in a good agreement with experimental data
up to $E_{\rm CM}\lsim 1.9$ GeV for the proton target.  In the 
higher energy region, $E_{\rm CM}\gsim 1.7$ GeV, note that $\rho$-meson
exchange interferes with various term.  As compared to the proton
case, the cross sections for the neutron target is underestimated in
the region of $E_{\rm CM}\gsim 1.7$ GeV.  We have tried to calculate the
differential cross sections with parameters varied in the
experimentally allowed region and found that it was possible to obtain
better results, compared to the GRAAL data.  However, it was not easy
at all to reproduce both the proton and neutron cases simultaneously in 
this higher energy region.  We expect that introducing other
background terms may help improve the results.  However, since we are
interested in understanding the newly found resonance $N^*$, not in
describing the data quantitatively well, we will not go further on to 
consider more resonances than here.  In the following, let us discuss
exclusively the properties of the peak structure at around $E_{\rm
  CM}\sim 1675$ MeV.      

\begin{figure}[t]
\begin{tabular}{cc}
\includegraphics[width=8cm]{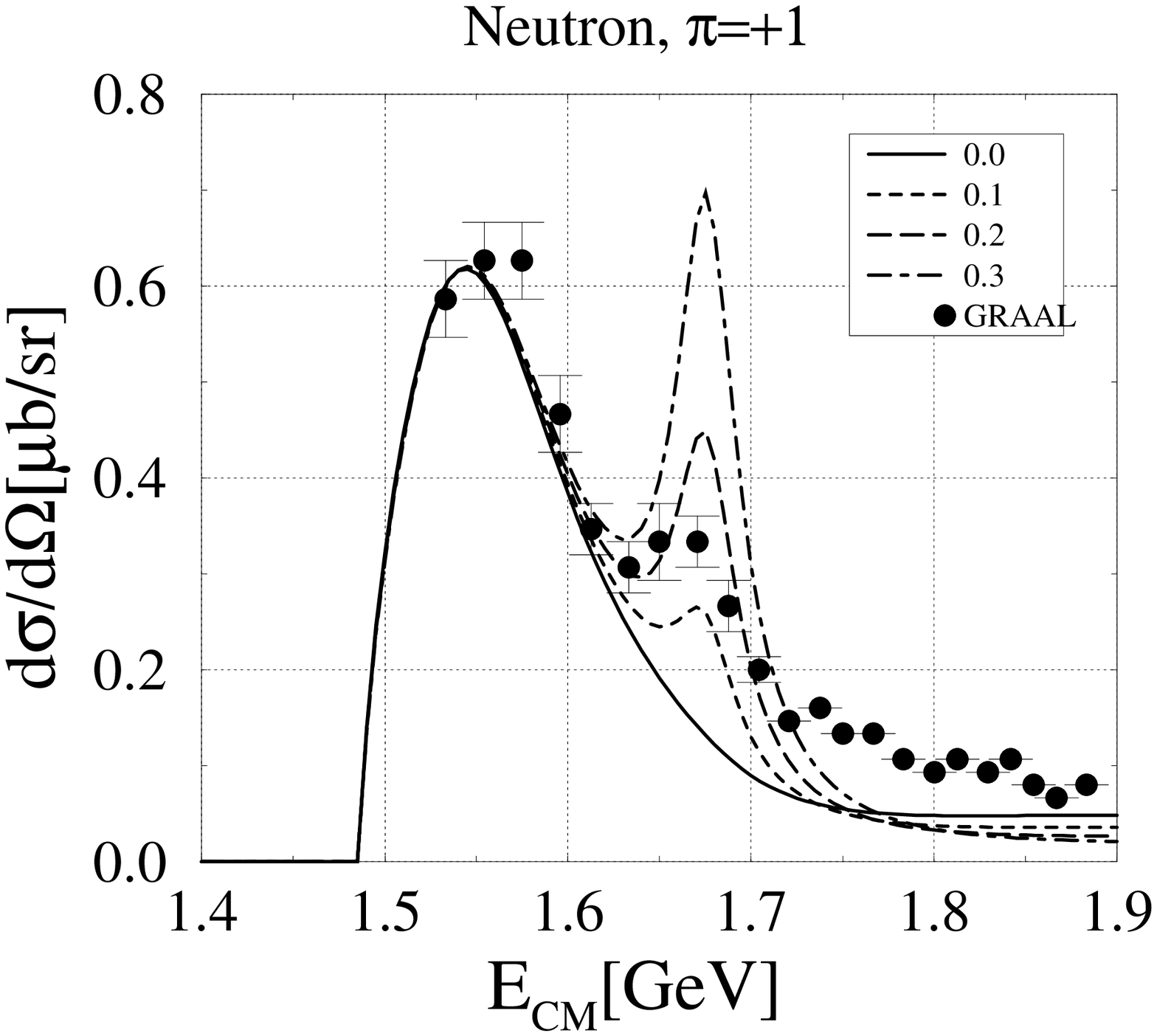}
\includegraphics[width=8cm]{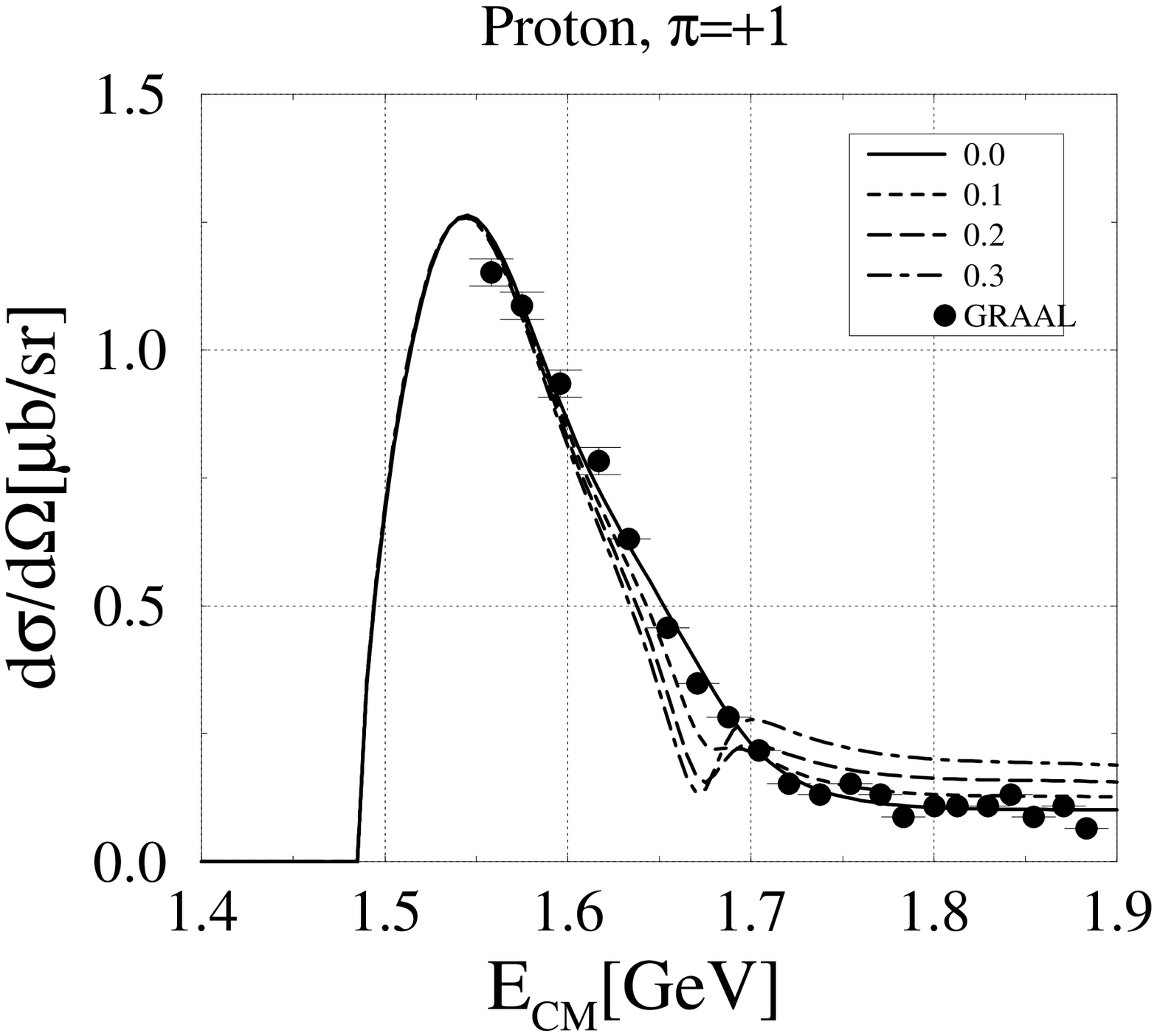}
\end{tabular}
\begin{tabular}{cc}
\includegraphics[width=8cm]{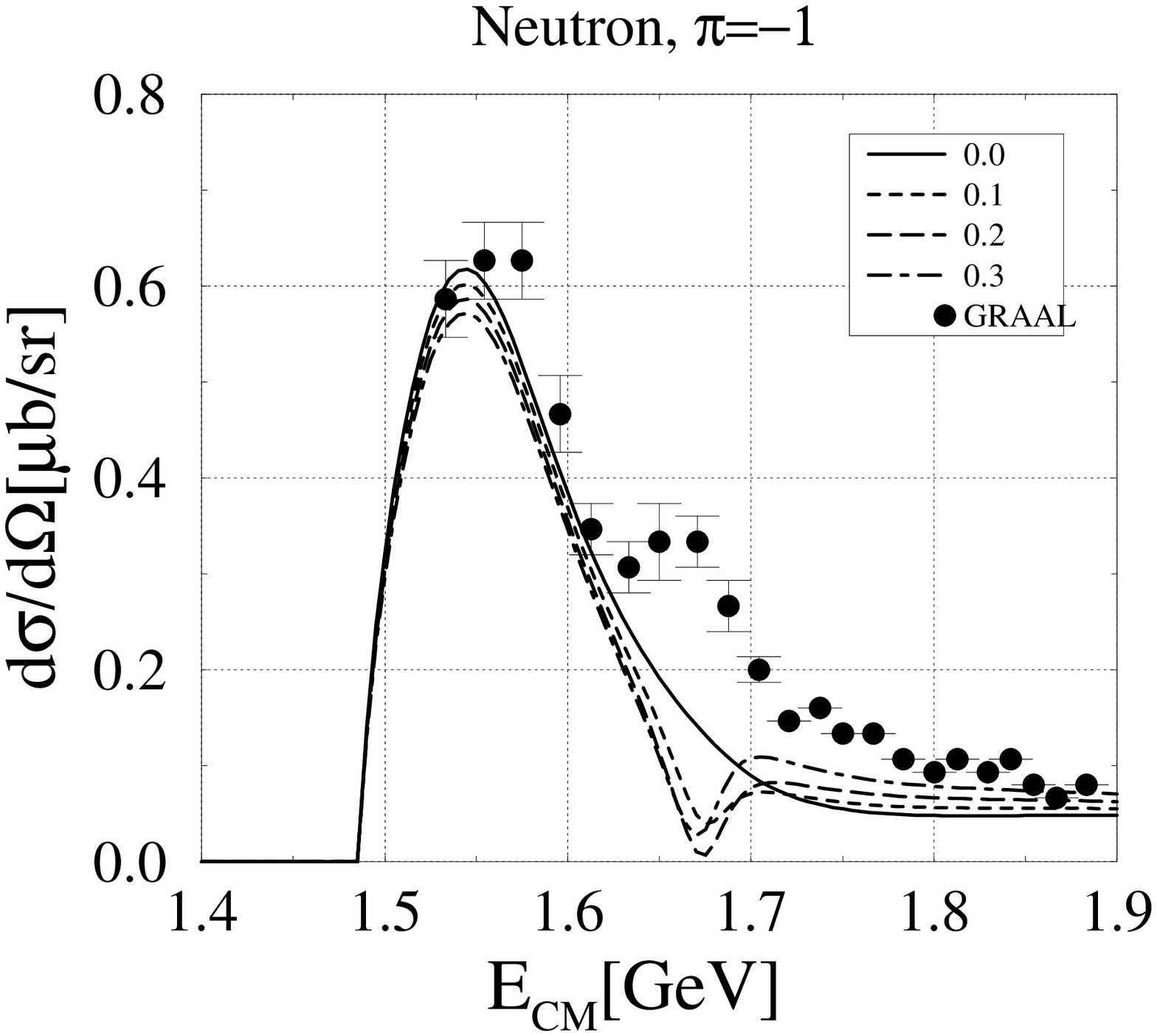}
\includegraphics[width=8cm]{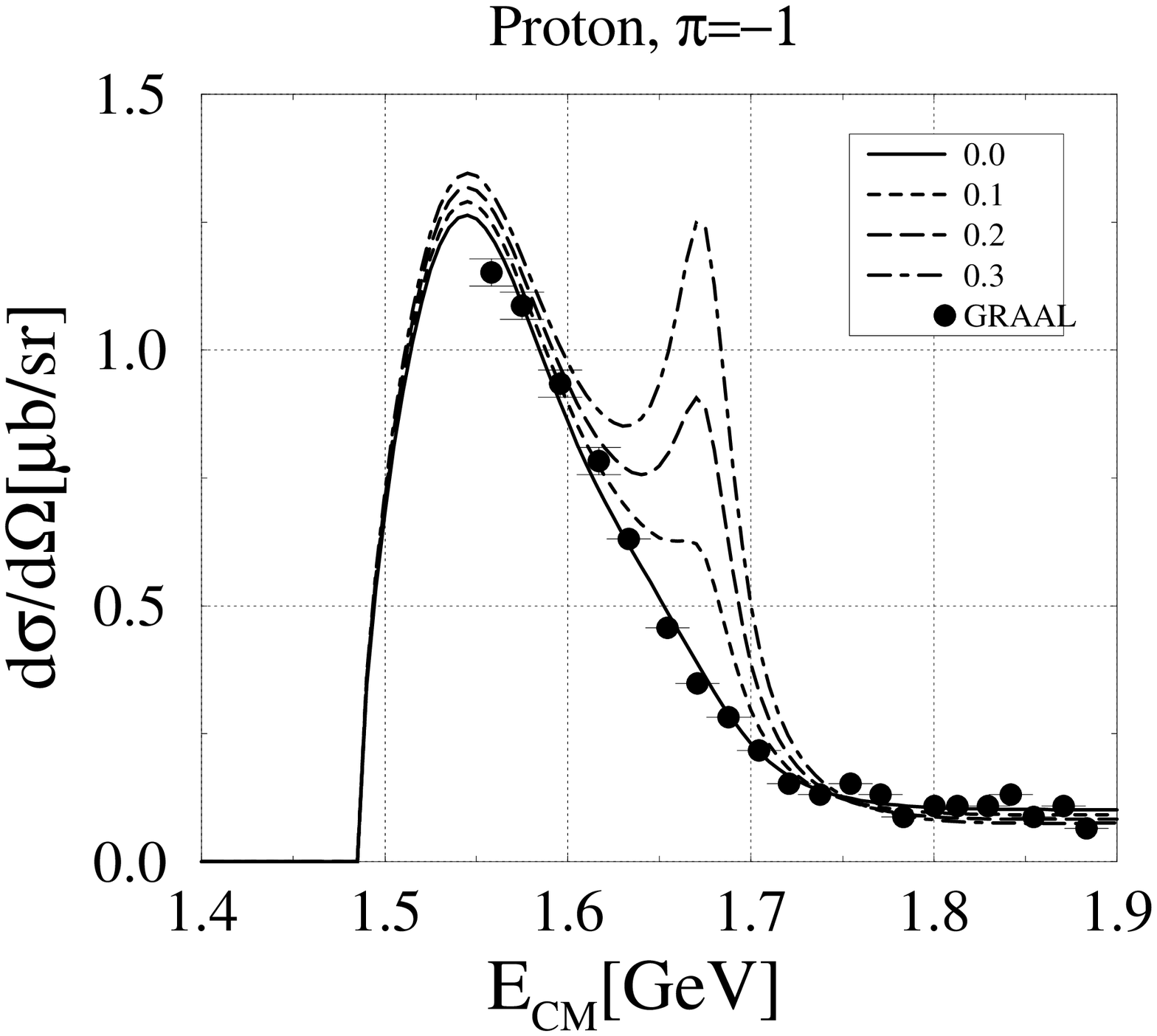}
\end{tabular}\caption{The differential cross sections as functions of
the total energy in the center of mass (CM) energy frame.  We depict
them in different targets (neutron at left column and proton at right
one), parities of $N^*(1675)$ (positive at upper two panels and
negative at lower two ones).  The four curves in each panel indicate 
$\mu_{\gamma NN^*}=0.0,\   0.1,\   0.2,\   0.3\, \mu_N$.  The
experimental data are taken from Ref.~\cite{Kuznetsov:2004gy}.}        
\label{fig2}
\end{figure}

In Fig.~\ref{fig2}, we show the results with the positive $\mu_{\gamma
NN^*}$.  With $\mu_{\gamma NN^*}$ turned on, the 
differential cross section starts to get changed around $E_{\rm CM}
\sim 1675$ MeV, showing various patterns of interference depending on
the parity of the $N^*$.  As $\mu_{\gamma NN^*}$ is
increased, two different patterns appear: While the differential cross 
sections of the $n$ ($p$) with positive (negative) parity show the
clearer peaks in the vicinity of $1675$ MeV, those of the $n$ ($p$)
with the negative (positive) parity are getting suppressed around that
energy.  These two different behaviors according to the parity of the
$N^*$ stem from either constructive or destructive interference among
the $N^*(1675)$ and other contributions.  In order to explain the
GRAAL data quantitatively with a positive $\mu_{\gamma
NN^*}$, we need to assume that the new nucleon-like resonance
should have positive parity.  Moreover, $\mu_{\gamma pp^*}$
must be much smaller than $\mu_{\gamma nn^*}$, which is
consistent with SU(3) flavor symmetry and the recent results of the
$\chi$QSM.  When $\mu_{\gamma nn^*}\;\simeq\; 0.2\, \mu_N$, the 
resonance structure is well described in the neutron channel.   

\begin{figure}[t]
\begin{tabular}{cc}
\includegraphics[width=8cm]{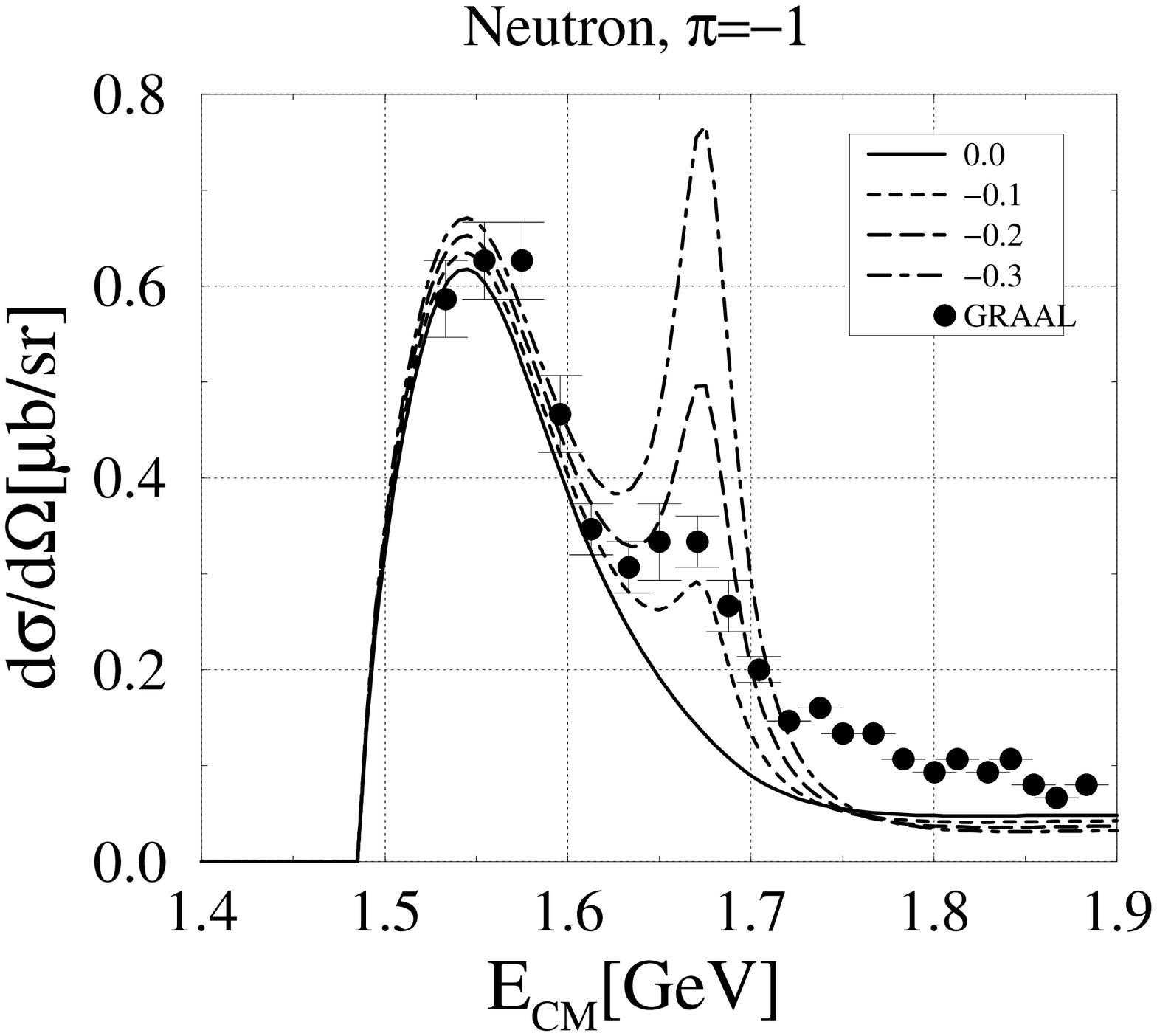}
\includegraphics[width=8cm]{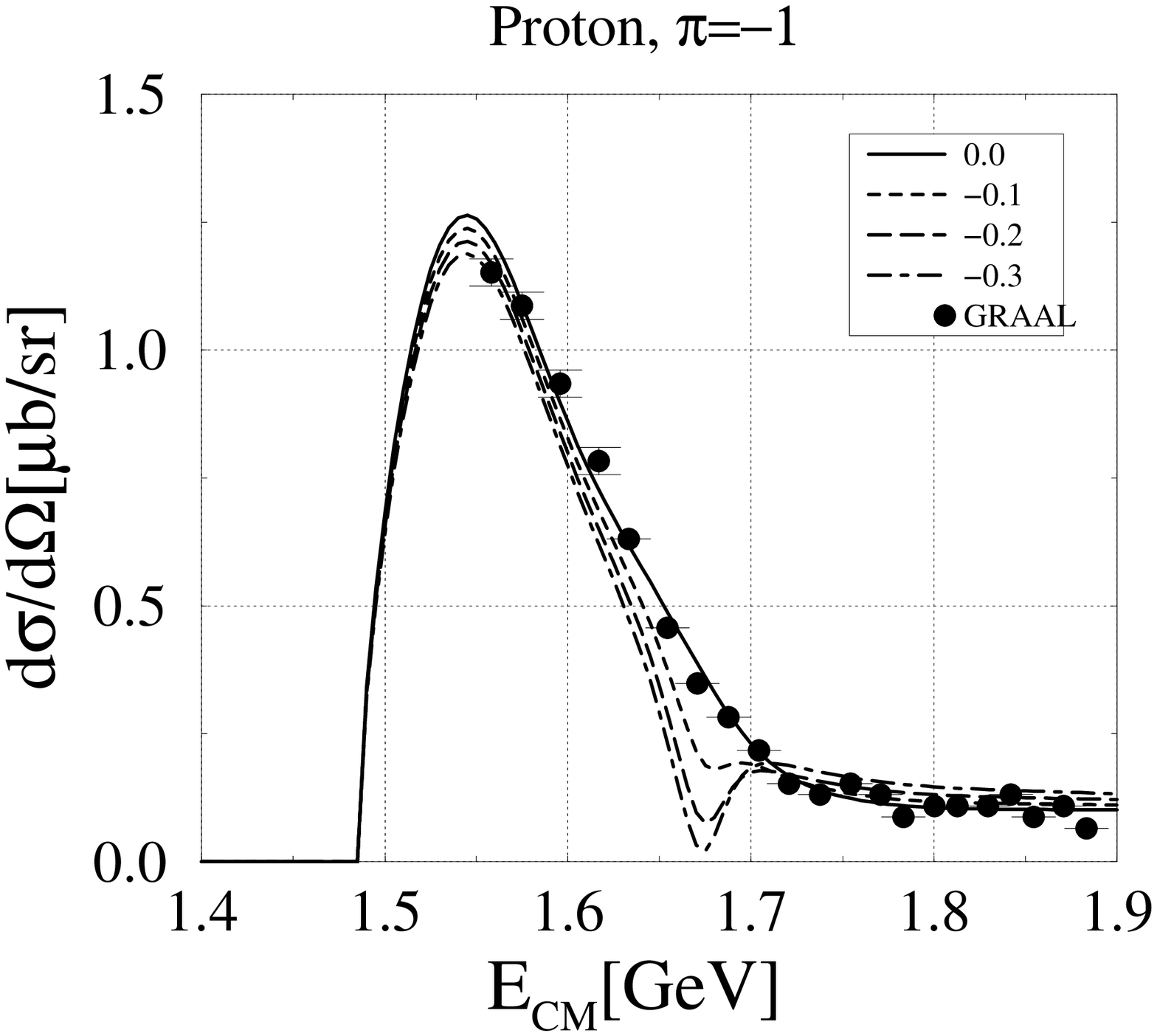}
\end{tabular}
\begin{tabular}{cc}
\includegraphics[width=8cm]{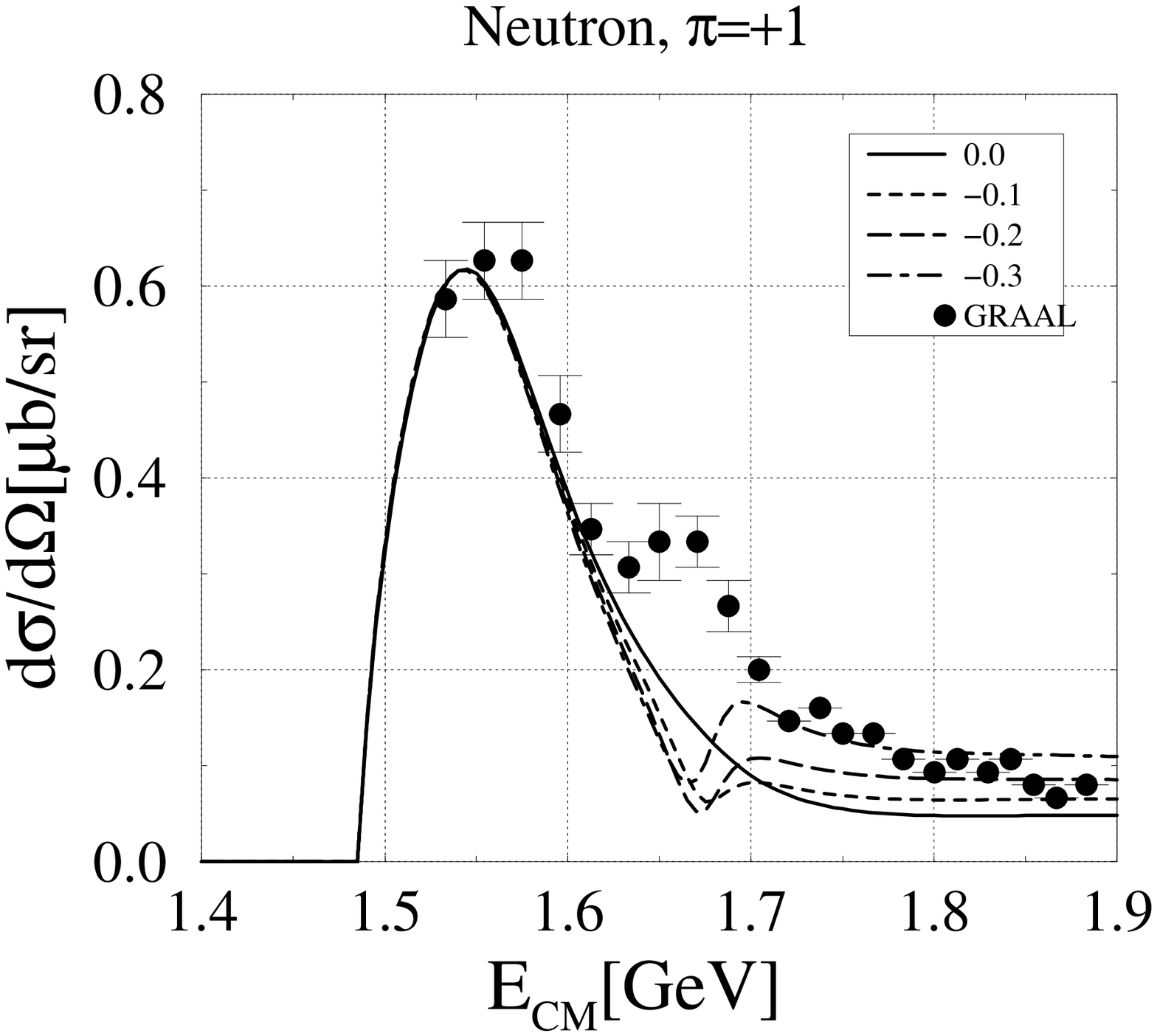}
\includegraphics[width=8cm]{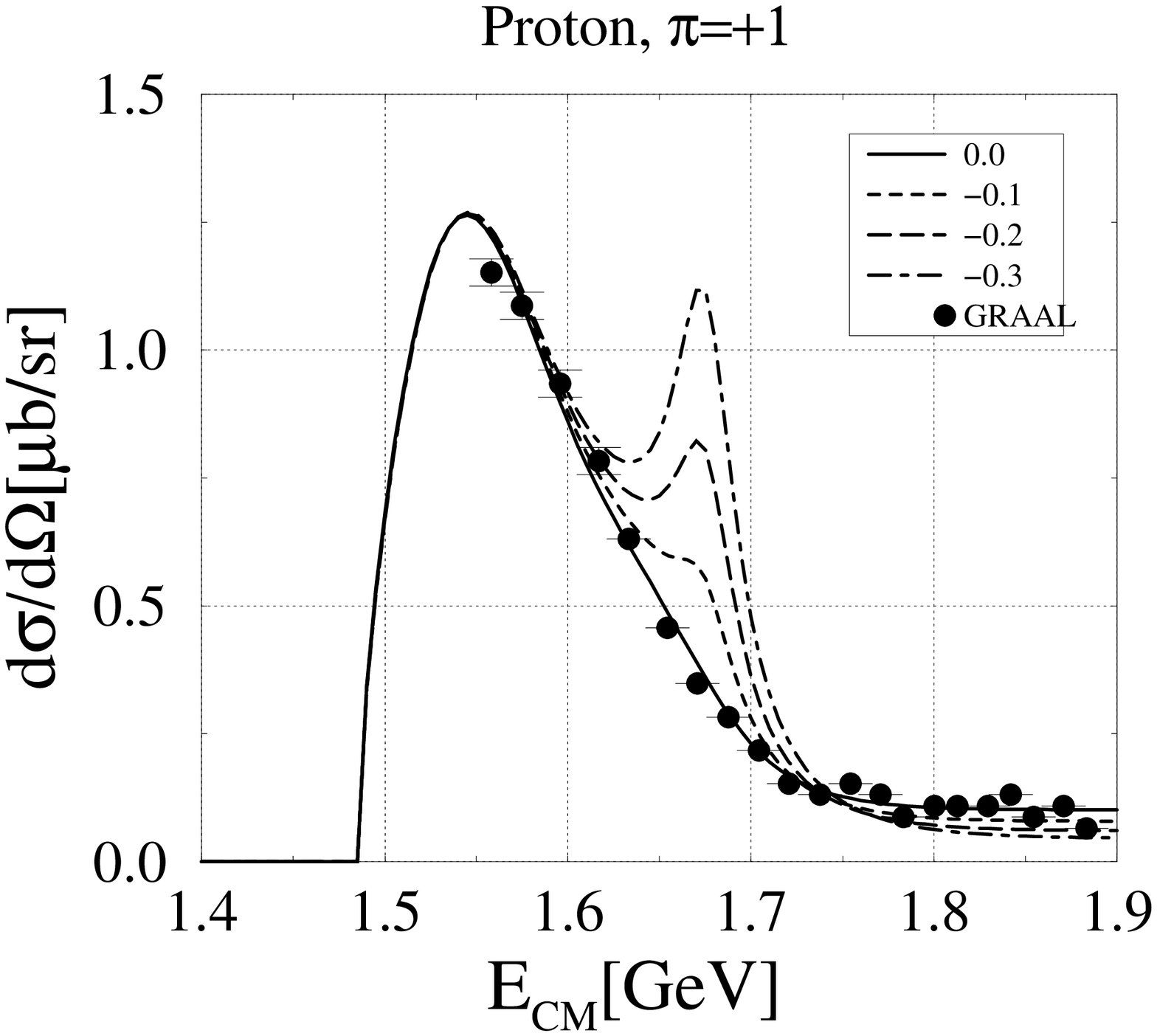}
\end{tabular}
\caption{The differential cross sections as functions of the total
energy in the center of mass (CM) energy frame.  We depict them in
different targets (neutron at left column and proton at left one),
parities of $N^*(1675)$ (positive at upper two panels and negative at
lower two ones).  The four curves in each panel indicate
$\mu_{\gamma NN^*(1675)}=0.0,\  -0.1,\  -0.2,\  -0.3\, \mu_N$.  The
experimental data are taken from Ref.~\cite{Kuznetsov:2004gy}.}    
\label{fig3}
\end{figure}

Fig.~\ref{fig3} shows the results with negative values of 
$\mu_{\gamma NN^*}$.  The tendency is similar to
Fig.~\ref{fig2} when the isospin of the target and parity are
interchanged simultaneously.  The peak structure of the neutron is
well reproduced when $\mu_{\gamma NN^*}\sim -0.2\, \mu_N$ as in
Fig.~\ref{fig3}.  However, in this case, the $n^*$ should have the
negative parity.  Thus, we conclude that the resonance structure of
the neutron may be explained by introducing the resonance $N^*(1675)$ 
with a finite magnetic transition couplings; $|\mu_{\gamma 
NN^*}|\sim 0.2\, \mu_N$.  This value is consistent with those
investigated in the $\chi$QSM and the phenomenological
study~\cite{Azimov:2005jj}. 

\section{Summary and Conclusion}
In the present work, we have investigated the $\eta$-photoproduction
via the reaction $\gamma N \to \eta N$, based on the effective
Lagrangians and the Born approximation.  Our focus was on the new
nucleon-like resonance $N^*(1675)$ observed in the recent GRAAL 
experiment~\cite{Kuznetsov:2004gy,Kuznetsov:2006de}.  We assumed that
$N^*(1675)$ was a pentaquark baryon identified as a member of the
baryon antidecuplet.  In order to make our study rather quantitative,
we included several nucleon resonances in addition to the nucleon pole
and vector meson exchanges.   Moreover, we included the new resonance
of $N^*(1675)$ with finite strengths of the electromagnetic
coupling constants $\mu_{\gamma N N^*}$.  

Since we do not know yet the parity of the resonance, we have
considered both positive and negative parities for the $N^*(1675)$.
The electromagnetic coupling is then the magnetic type for the positive
parity case and electric type for the negative parity one.  In both
cases,  we were able to describe the GRAAL data well, using the 
transition magnetic moments $|\mu_{\gamma nn^*(1675)}|\sim 0.2\, \mu_N$ and 
$|\mu_{\gamma pp^*(1675)}|\sim 0$.  It implies that it is quite 
reasonable to regard the new nucleon-like resonance in the GRAAL data
as a nucleon partner of the pentaquark baryon belonging to the SU(3)
antidecuplet.   

\section*{Acknowledgement}
This work was supported by the Korea Research Foundation Grant funded
by the Korean Government(MOEHRD) (KRF-2005-202-C00102).  The work of
KSC was supported in part by the Japan Student Services Organization
due to the program, "Osaka University Regular student exchange course
2004-2005".  The work of A.H. was supported in part by the Grant for
Scientific Research ((C) No.16540252) from the Ministry of Education,
Culture, Science and Technology of Japan.   
 
\end{document}